\documentclass[12pt,epsfig]{article}

\renewcommand{\baselinestretch}{1.2}

\usepackage[dvips]{graphics}
\usepackage{rotating}
\usepackage{epsfig}
\def\lesssim{\mathrel{\hbox{\rlap{\hbox{\lower5pt\hbox{$\sim$}}}\hbox{$<$}}}}
\def\gtrsim{\mathrel{\hbox{\rlap{\hbox{\lower5pt\hbox{$\sim$}}}\hbox{$>$}}}}

\newcommand{\ntrl}[1]{\chi^0_#1}

\newcommand{\chp}[1]{\chi^+_#1}
\newcommand{\chm}[1]{\chi^-_#1}
\newcommand{\sbot}[1]{\tilde{b}_#1}
\newcommand{\sstop}[1]{\tilde{t}_#1}

\newcommand{\sbotc}[1]{\tilde{b}_#1^*}
\newcommand{\sstopc}[1]{\tilde{t}_#1^*}
\def\gluino{\tilde{g}}

\newcommand{\mntrl}[1]{m_{\chi^0_#1}}      %
\newcommand{\mchpm}[1]{m_{\chi^\pm_#1}}

\newcommand{\msbot}[1]{m_{\tilde{b}_#1}}
\newcommand{\mstop}[1]{m_{\tilde{t}_#1}}
\def\mgluino{m_{\tilde{g}}}

\def\tanbeta{\tan\beta}

\def\ptmiss{\not\!\!{p_T}}

\def\beq{\begin{equation}}   %
\def\eeq{\end{equation}}   %

\begin{document}

\begin{flushright}
   {\bf KIAS-P10024}
\end{flushright}

\vskip 30pt

\begin{center}
{\large \bf  Higgs production in CP-violating supersymmetric cascade decays:
probing the `open hole' at the Large Hadron Collider}\\
\vskip 20pt
{Priyotosh Bandyopadhyay$^{a}$\footnote{priyotosh@kias.re.kr}}
  \\
\vskip 20pt
{ Korea Institute for Advanced Study \\
 Hoegiro 87(207-43 Cheongnyangni-dong),\\
 Dongdaemun-gu, Seoul 130-722, Korea}\\

\end{center}

\vskip 65pt

\abstract{A benchmark $CP$-violating supersymmetric scenario (known as `CPX-scenario' in the literature)
is studied in the context of  the Large Hadron Collider (LHC). It is shown
that the LHC, with low to moderate
accumulated luminosity, will be able to probe the existing `hole' in the 
$m_{h_1}$-$\tan\beta$ plane, which cannot be ruled out by the LEP data. We explore the parameter space with cascade decay 
of third generation squarks and gluino with CP-violating decay branching fractions. We propose a multi-channel analysis to probe this parameter space some of which are background free at an integrated luminosity of 5-10 fb$^{-1}$. Specially, multi-lepton final states ($3\l,\, 4\l$ and like sign di-lepton)
are almost background free and have $5\sigma$ reach for the corresponding signals with very early data of LHC for both 14 TeV and 7 TeV center of mass energy.
}

\newpage

\section{Introduction}

One of the main motivations for suggesting supersymmetry (SUSY) is to remove the fine-tuning
problem in the Higgs sector of the standard model.
The condition of holomorphicity of the superpotential requires two Higgs doublets
in the minimal SUSY extension of the standard model (SM). There the Higgs sector has a 
larger particle content than the SM, and the physical states in this sector
comprise two neutral scalars, one pseudoscalar and one charged Higgs boson.
Finding the signatures of these  scalars is also important along with
the search for SUSY at the upcoming Large Hadron Collider (LHC).

Prior to the LHC several experiments failed to discover the Higgs, but yielded some bound on the Higgs mass. The strongest
lower bound on the smallest Higgs mass ($m_h$)  from the Large Electron Positron Collider (LEP) 
is $m_h > $ 114.4 GeV \cite{Barate:2003sz,Schael:2006cr}. In the MSSM, with all the real and CP-conserving parameters, the lower limit on the lightest Higgs boson is $\sim$ 90 GeV \cite{susylim} for any $\tan{\beta}$. However, when the Higgs sector inherits some
$CP$-violating phase through radiative corrections \cite{Pilaftsis:1998pe,Pilaftsis:1998dd}, the above limit ceases to be valid. We will be concentrating in this CP-violating scenario.

CP-violation in the Higgs sector is possible in multi-Higgs doublet models, 
such as a general 2-Higgs Doublet Model (2HDM) or indeed the MSSM. 
In the latter, it has been shown that, assuming universality of the 
gaugino masses $(M_i, i =1,2,3)$ at some high energy scale, the CP-violating
MSSM Higgs sector can be parametrised in terms of two independent phases: 
that of the Higgsino mass parameter (also called  $\mu$ term), i.e.,
 ${\rm Arg}(\mu)$, 
and that of the soft trilinear Supersymmetry (SUSY) 
breaking parameters, i.e., ${\rm Arg}(A_f) $, with $f=t,b$. 
The experimental upper bounds on the Electric Dipole Moments (EDMs) of
 electrons and  neutrons \cite {edm1, edm2} as well as of mercury
 atoms \cite{edm3} constraints on these phases. 

It is well-known by now that the lower bound on the mass of the lightest Higgs boson of 
the $CP$-conserving MSSM (from LEP \cite{Schael:2006cr}) can be drastically reduced or may even 
entirely vanish if non-zero 
$CP$-violating phases are allowed \cite{Bechtle:2006iw,Carena:2000ks}. This can happen through radiative corrections to the Higgs potential, whereby the above mentioned phases of the Higgsino mass parameter $\mu$ 
and the trilinear soft SUSY breaking parameter $A$ enter into the picture. As a result 
of the $CP$-violating phase, the neutral spinless states are no more of definite parity,
and their couplings to gauge bosons as well as fermions are thus modified, depending on
the magnitude of the phases. Thus there are three neutral states $h_i$ ($i$=1,2,3);
the collider search limits for all of them are modified since the squared amplitudes
for production via $WW$, $ZZ$ and $q\bar{q}$ couplings for all of them now consist of
more than one term. Due to this mixing through the loop effects, the lightest Higgs boson is almost CP-odd with highly suppressed coupling to $ZZ$ pair. Thus results in reduced production rates and consequent weakening of mass limits at collider experiments.

In the context of a
benchmark $CP$-violating scenario (often called the CPX scenario in the
literature \cite{Carena:2000ks}), 
it has been found that $m_{h_1}$ as low as  50 GeV or even smaller, cannot be ruled out by the 
final LEP data for low and moderate values of  $\tan\beta$,\footnote{where $h_1$ is the lightest
neutral Higgs, and $\tan\beta$ is the 
ratio of the vacuum expectation values of the two Higgs doublets}. 
In other words, a `hole' is found to exist in the $m_{h_1}$-$\tanbeta$ parameter space 
covered by the LEP searches, the underlying reason being the reduction
in  the coupling $ZZh_1$ due to the $CP$-violating phase(s), as mentioned above. 
Moreover, complementary channels such as
$e^+ e^- \to h_1 h_2$,  suffer from coupling as well as phase-space suppression 
within this `hole', thus making it inaccessible to  LEP  searches.
The existence of this hole has been confirmed by the analysis of 
the LEP data by different experimental groups \cite{Schael:2006cr}, although
its span varies.

The next natural step is to assess the prospect of closing the hole at Tevatron 
Run II or the LHC. 
The existing analysis on this \cite{Carena:2002bb}, however, focuses on the discovery channels based on
the conventional Higgs production and decay 
mechanisms  employed in the context of the SM.  It has been noted 
that although the hadron colliders can probe most of the parameter space of the 
CPX scenario and can indeed go beyond some regions of the parameter space scanned
by the  LEP searches, the lightest Higgs boson within the aforementioned hole may 
still escape detection. This is because not only the $ZZh_1$ but also the $WWh_1$ and 
$t\bar{t}h_1$ couplings tend to be very small within this hole.  
On the other hand, the relatively heavy neutral Higgs bosons $h_{2,3}$ couple to $W$, $Z$ 
and $t$ favourably, but they can decay in non-standard channels, thus requiring
a modification in search strategies.
The other work \cite{Accomando:2006ga} which has looked into possible signals of the
CPX scenario at the LHC is also restricted  to the
production of $h_i$ ($i$=1,2,3) bosons in SM-like channels. However, they
looked into more decay channels of the $h_i$ bosons thus produced.
Until now it has been concluded that parts of
the holes in the $M_H^+$-$\tan\beta$ or the $m_{h_1}$-$\tan\beta$
parameter space can be plugged, although considerable portions of the hole, 
especially for low $\tan\beta$, may escape detection
at the  LHC even after accumulating 300 fb$^{-1}$ of integrated luminosity.

Thus it is important to look for other production channels for the scalars in the 
CPX region, especially by making use of the couplings of $h_1$ with the sparticles. In this context we explore the cascade decay of third generation scalar quarks, mainly $\tilde{t}_{1}\tilde{t}^{*}_{1}$ and $\tilde{b}_{1}\tilde{b}^{*}_{1}$.
This could be discovery channels, in cases where the $t$-$\bar{t}$-$h_1$ 
and  $W$-$W$-$h_1$, $Z$-$Z$-$h_1$  couplings are highly suppressed. 

Previously, measurement of CP-asymmetries
and the various decays probing this CP-violating sector have been studied extensively [\cite{Ghosh:2004cc}-\cite{Das:2010ds}].
The Higgs production under the CP-conserving supersymmetric cascade has been analysed
in Refs. \cite{Datta:2001qs,Datta:2003iz,Bandyopadhyay:2008fp,Bandyopadhyay:2008sd,Kribs:2009yh}.
It has been noted that in a general $CP$-violating MSSM, 
the cross section of $\sstop1\sstopc1h_1$ production 
could be dramatically larger than that obtained by switching off the $CP$-violating 
phases \cite{Li:2006hq}. Since the trilinear SUSY breaking parameter $A_t$ 
is necessarily large in the CPX scenario, 
$\sstop1$ tends to be relatively light and may be produced at 
the LHC with large cross section. As a bonus, both $h_2$ and $h_3$ also couple 
favourably to the $t\bar{t}$ pair and can add modestly to the signal although by 
themselves they fail to produce a statistically significant signal. In Ref. \cite{Bandyopadhyay:2007cp} we investigated
the implications of these couplings at the LHC, by concentrating on a specific signal 
arising from the associated production of the neutral Higgs bosons with a 
top-pair or a pair of lighter stop squarks.

This paper is organised as follows. In Section 2 we discuss the basic
inputs of the CPX scenario, the resulting mass spectrum and other
features they lead to. All of our subsequent numerical analysis would be
in this framework where we also use the alternative expression CPV-SUSY to 
mean the CPX-scenario.  In section 3 we define the proposed parton level
 signal. In section 4 we do the collider simulation and devise the event
 selection criteria to reduce the SM backgrounds and present the final
 numerical results in section 5. We summarise and conclude in section 6.

\section{The CPX  Model: values of various parameters} 

As indicated in the introduction, we adopt the so called CPX 
scenario in which the LEP analyses have been
performed. It has been observed
\cite{Pilaftsis:1998pe, Pilaftsis:1998dd}
that the $CP$-violating quantum effects on the Higgs potential is proportional to
$Im(\mu A_t)/M^2_{SUSY}$, where $A_t$ is the trilinear soft SUSY breaking parameter occurring
in the top squark mass matrix, and $M_{SUSY}$ is the characteristic SUSY breaking scale,
which is of the order of the third generation squark masses. 
With this in mind, a benchmark scenario known as CPX was proposed and
its consequences were studied
\cite{Pilaftsis:1999qt, Demir:1999hj, Choi:2000wz, Kane:2000aq, Choi:2001pg, Choi:2001iu,
 Heinemeyer:2001qd, Ibrahim:2002zk,  Ham:2002ps, twohiggs}.
In this scenario, the effects of $CP$-violation are maximized.
The corresponding inputs that we adopt here 
are compatible with the ``hole'' left out in the analysis.

\hskip 47pt $m_{\tilde{t}} = m_{\tilde{b}} = m_{\tilde {\tau}} = M_{SUSY} = 500$ GeV, 
 $\qquad \mu = 4 M_{SUSY} = 2$ TeV 

\hskip 50pt $|A_t| = |A_b| = 2 M_{SUSY} = 1$ TeV, $\quad arg(A_{t,b}) = 90^\circ$ 

\hskip 50pt $|\mgluino|= 1$ TeV, $\quad arg(\mgluino) = 90^\circ$

\hskip 50pt  $M_2 = 2 M_1 = 200$ GeV,~~~~$\tan\beta= 5 - 10$ \\

\noindent
The only departure from reference \cite{Carena:2002bb} lies in a small tweaking 
in the mass ratio of the $U(1)$ and $SU(2)$ gaugino masses 
$M_1$ and $M_2$, aimed at ensuring gaugino mass unification at high scale. 
It has been checked that this difference does not affect the Higgs production
or the decay rates.
The presence of a relatively large $A_t$ ensures that one of the top squarks will
be relatively light. The value of the top quark mass has been taken to be 175 GeV\footnote{The frequent shift in the central
  value of $m_t$, coming from Tevatron measurements, causes the size of the
  hole to change, although its location remains the same. 
However, there is little point in worrying about this uncertainty, since the 
very quantum corrections which are at the root of all $CP$-violating 
effects in the Higgs sector are prone to similar, if not greater, 
theoretical uncertainties}.

It is to be noted that the first two generation sfermion masses must be kept
sufficiently heavy so that the stringent experimental bound (for example, 
the electric dipole moment of the neutron) is satisfied. Here we have not 
considered possible ways of bypassing such bounds, and set the masses of the first 
two sfermion families at 10 TeV. Thus our analysis is based on
the mass spectrum showed in Table \ref{tabmass} with $\tan\beta$=5 and  $m_{H^\pm}$=$130$ GeV with the threshold corrections and considered as benchmark point 1(BP1).The cross sections for different supersymmetric production processes are computed with {\tt CalcHEP} \cite{Pukhov:2004ca} (interfaced with the program {\tt CPSuperH}\cite{Ellis:2006eh,Lee:2003nta}). For the benchmark point 1(BP1) the cross-section for different supersymmetric processes are listed in the Table \ref{tabcross1}. The cross-sections of the associated Higgs production within the model have been given in Table \ref{crossH}.

\begin{table}
\begin{center}
\renewcommand{\arraystretch}{1.4}
\begin{tabular}{|c|c|c|c|c|c|c|c|c|c|c|}
\hline
$m_{h_1}$ & $m_{h_2}$ & $m_{h_3}$ & $\mstop1$ & $\mstop2$ & $\msbot1$
& $\msbot2$ & $\mntrl1$ & $\mntrl2$ & $\mchpm1$ \\
\hline
39.8 & 104.7 & 137.1 & 317.6 & 668.2 & 475.9 & 526.6 & 99.6 & 198.4 & 198.4 \\
\hline
\end{tabular}
\caption{Physical masses (in GeV) of neutral Higgs bosons, squarks and lighter
  gauginos in the CPX scenario with $\tan\beta$=5 and  $m_{H^\pm}$=$130$ GeV with the threshold corrections.}\label{tabmass}
\end{center}
\end{table}

 \vskip 10pt
 \renewcommand{\arraystretch}{1.4}
\begin{table}
 \begin{center}
\begin{tabular}{|c|c|c|c||c|c|c||c|}
 \hline
 $\sigma_{\sstop1 \sstopc1 }$  &  $\sigma_{\sbot1 \sbotc1 }$  & $\sigma_{\sstop2 \sstopc2 }$  & $\sigma_{\sbot2 \sbotc2 }$ &  $\sigma_{\sstop1 \sstop2 }$  &  $\sigma_{\sbot1 \sbot2 }$  & $\sigma_{\tilde{t}_i \tilde{b}_j }$ &$\sigma_{ \tilde{g}\tilde{g}}$ \\
\hline
 2861 & 323.3 & 4 &178.5 & 8&0.6&7 &135\\
 \hline
\end{tabular}
 \caption{Production cross sections (in fb) at lowest-order 
computed with {\tt CalcHEP} interfaced with CPsuperH
 for different signal processes at the LHC in the
CPX scenario and for the spectrum of Table 1. 
 {\tt CTEQ6L} parton distribution functions are used and
 the renormalization/factorization scale is set to  $\sqrt{\hat{s}}$.}\label{tabcross1}
\end{center}
 \end{table}

 
\renewcommand{\arraystretch}{1.0}
\begin{table}

\begin{center}
\begin{tabular}{||c|c|c||}
 \hline
$\sigma_{t \bar{t} h_1}$  &  $\sigma_{t \bar{t} h_2}$  & $\sigma_{t \bar{t} h_3}$\\
\hline
8 &190 &132\\
 \hline
\end{tabular}
 \caption{Production cross sections (in fb) at lowest-order 
computed with {\tt CalcHEP} interfaced with CPsuperH
 for different signal processes at the LHC in the
CPX scenario and for the spectrum of Table 1. 
 {\tt CTEQ6L} parton distribution functions are used and
 the renormalization/factorization scale is set to  $\sqrt{\hat{s}}$.}\label{crossH}
\end{center}
\end{table}

As we will be focusing on the cascade decay of the third generation strongly interacting supersymmetric particles (squarks)  and gluino we should also see how the branching fraction changes down the cascade compared to CP-conserving SUSY scenario. For the bench mark point we list the the branching fractions for $\sstop1$, $\sbot1$, $\tilde{g}$ and the Higgses in the Tables \ref{brstop1}, \ref{brsbot1}, \ref{brgluino}, \ref{brHiggses} respectively. Here the charged Higgs which is produced in the cascade decay goes through an unconventional decay mode $H^{\pm} \to h_1 W^{\pm}$ due to the low $m_{h_1}$ in CPX scenario. Mainly due to this all the signal topologies get changed that is from the CP-conserving SUSY scenario. 


 \renewcommand{\arraystretch}{1.0}
\begin{table}
 \begin{center}
\begin{tabular}{||c|c||}
 \hline
Br($\sstop1 \to b \chp1$) &  Br($\sstop1 \to t \ntrl1$) \\
 \hline
0.81 & 0.19 \\
 \hline
\end{tabular}
 \end{center}
\caption{Branching fractions for lighter top squark in the CPX scenario.}\label{brstop1}
\end{table}


\renewcommand{\arraystretch}{1.4}
\begin{table}
 \begin{center}
\begin{tabular}{||c|c|c|c|c||}
 \hline
Br($\sbot1 \to \sstop1 H^-$)&Br($\sbot1 \to \sstop1 W^-$) & Br($\sbot1 \to t \chm1$)& Br($\sbot1 \to b \ntrl2$)& Br($\sbot1 \to b \ntrl1$) \\
 \hline
0.77& 0.12 &  0.05& 0.04 &0.01 \\
 \hline
\end{tabular}
 \end{center}
\caption{Branching fractions for lighter bottom squark in the CPX scenario.}\label{brsbot1}
\end{table}



\begin{table}
 \begin{center}
\begin{tabular}{||c|c|c|c||}
 \hline
Br($\gluino \to b \sbot1$)&Br($\gluino \to b \sbot2$) & Br($\gluino \to t \sstop1$)& Br($\gluino \to t \sstop2$) \\
 \hline
 0.28 &  0.24& 0.32 &0.16 \\
 \hline
\end{tabular}
 \end{center}
\caption{Branching fractions for gluino in the CPX scenario.}\label{brgluino}
\end{table}


\begin{table}
 \begin{center}
\begin{tabular}{||c|c|c|c||}
 \hline
 Br($H^{\pm} \to h_1 W^{\pm}$)& Br($h_1 \to b \bar{b}$)  & 
 Br($h_2 \to h_1 h_1 $) &Br($h_3 \to h_1 h_1$)   \\
 \hline
0.84&0.92 & 0.85 & 0.82   \\
 \hline
\end{tabular}
 \end{center}
\caption{Branching fractions for Higgs bosons in the CPX scenario.}\label{brHiggses}
\end{table}


\section{Collider signatures}
In this section we will discuss the cascade decays of third generation squarks  
under the CPX scenario. First consider the $\sstop1 \sstopc1$. When both the $\tilde{t_1}$ decays via $\tilde{t_1}\to t \ntrl1$ then we have 
the following final states at the end.

$$
pp \to \tilde{t_1}\tilde{t_1}^* \to t\bar{t}\ntrl1\ntrl1 \to 
 b\bar{b}H^+H^-\ntrl1\ntrl1 
$$
$$
\to  b\bar{b}W^+W^-h_1h_1\ntrl1\ntrl1
$$
$$
\to 6b + 4(non- b)jet+ \not{p_T}
$$
$$
\to 6b +l + 2(non- b)jet+ \not{p_T}
$$
$$
\to 6b +OSD + \not{p_T}
$$
But Br$(t \to b H^+)\simeq 0.011$ due to which the effective branching 
fraction, i.e. Br$(\sstop1 \sstopc1 \to b\bar{b}H^+H^-\ntrl1\ntrl1) \simeq 5\times 10^{-8}$. Thus above channels which could have been  interesting and could have added to the signal cross-section those coming from $\tilde{b_1} \tilde{b_1}^*$ no longer do so. There are other modes coming from $\sstop1 \sstopc1$ which could be interesting. For example, the case where one of the $\tilde{t_1}$ decays via $\sstop1 \to b \chp1$ and this gives rise to the following signal signal topologies.

$$
\tilde{t_1}\tilde{t_1}^* \to t\bar{b}\ntrl1\chm1 \to  b\bar{b}H^+W^-\ntrl1\ntrl1 \to b\bar{b}h_1W^+W^-\ntrl1\ntrl1
$$
$$
\to 4b+4(non- b)jet+ \not{p_T}
$$
$$
\to 4b+1(non- b)jet+ 1\ell +\not{p_T}
$$
$$
\to 4b+ OSD+ \not{p_T}
$$

Now we consider the other pair production of the third generation scalar quark, the sbottom pair production. As given in Table \ref{tabcross1} s the production cross section is about 323 fb. Due to large branching fraction in $\tilde{b_1}\to\tilde{t_1} H^-$ (Table \ref{brsbot1}), in the main decay mode, both of the sbottoms decay in this channel. Then due to large branching fraction of $H^{\pm} \to h_1 W^{\pm}$ (Table \ref{brHiggses}) we end up with $4 W+ 2 h_1+ 2 b + \ptmiss$. Depending on the decay mode of $w$ we can have the following final states.

$$
pp \to \tilde{b_1}\tilde{b_1}^* \to \tilde{t_1}\tilde{t_1}^*H^+H^- \to 
 b\bar{b}W^+W^-W^+W^-h_1h_1+ \not{p_T} 
$$
$$
\to 6b +LSD + 4(non- b)jet+ \not{p_T}
$$
$$
\to 6b +3\ell + 2(non- b)jet+ \not{p_T}
$$
$$
\to 6b +4\ell + \not{p_T}
$$

The contribution of gluino pair production comes from the gluino decays to these
stops and sbottoms (Table \ref{tabcross1} and Table \ref{brgluino}). As for this CPX parameter space, mass of the gluino is taken to be heavier than the third generation scalar quarks, so this cross-section just adds to the signal cross-section (either stop or sbottom). Below we list the effective cross-section of different channels coming from third generation scalar squark decays (stop and sbottom pairs).


\renewcommand{\arraystretch}{1.0}
\begin{table}
\begin{center}
\begin{tabular}{||c|c|c||}
 \hline
Number of & Channels& Effective\\
channels  & & cross-sec (in fb)\\
\hline
\hline
1 &$ 6b +LSD + 4(non- b)jet+ \not{p_T}$ & 11.49 \\
 \hline
1 &$ 6b +OSD + 4(non- b)jet+ \not{p_T}$ & 22.98 \\
 \hline
2&$ 6b +3\ell + 2(non- b)jet+ \not{p_T}$&17.24 \\
\hline
3 &$6b +4\ell + \not{p_T}$& 8.62\\
\hline
4&$4b+4(non- b)jet+ \not{p_T}$&0.38\\
\hline
5&$4b+1(non- b)jet+ 1\ell +\not{p_T}$&0.18\\
\hline
6&$4b+ OSD+ \not{p_T}$&0.09\\
\hline
\end{tabular}
 \caption{Production cross sections (in fb) at lowest-order 
computed with {\tt CalcHEP} interfaced with CPsuperH
 for different signal processes at the LHC in the
CPX scenario and for the spectrum of Table 1. 
 {\tt CTEQ6L} parton distribution functions are used and
 the renormalization/factorization scale is set to  $\sqrt{\hat{s}}$.}\label{cross-sec}
\end{center}
\end{table}


The channels coming from $\tilde{b_1}\tilde{b_1}^*$ are absolutely
background free at the partonic level. So these could be golden channels
 to probe CPX. It is the $\tilde{b_1}\tilde{b_1}^*$ which could produce 
multiple $b$-s in the final state. It is very clear from Table 
\ref{cross-sec} that the multiple $b$ processes, like the $4b$ processes
 coming from $\tilde{t_1}\tilde{t_1}^*$ have very low cross-section.
But because of ISR/FSR and jet smearing, the parton level predictions
 could change. For that purpose we go through a PYTHIA \cite{Sjostrand:2001yu} 
 simulation to analyse the signals and backgrounds in the next section. 

\section{Collider study}

In this study, {\tt CalcHEP} (interfaced to the program {\tt CPSuperH}) has also been used for generating parton-level events for the relevant processes. The standard {\tt CalcHEP-PYTHIA} interface \cite{alex}, which uses the SLHA
 interface \cite{Skands:2003cj} was then used to pass the {\tt CalcHEP}-generated 
events to {\tt PYTHIA} \cite{Sjostrand:2001yu}. Further, all relevant decay-information is generated with 
 {\tt CalcHEP} and is passed to {\tt PYTHIA} through the same interface.  All these 
 are required since there is no public implementation of CPV-MSSM in {\tt PYTHIA}. Subsequent 
decays of the produced particles, hadronization and the collider analyses are done with {\tt PYTHIA (version 6.4.22)}. 

We used {\tt CTEQ6L} parton distribution function (PDF) \cite{Lai:1999wy,Pumplin:2002vw}. 
In {\tt CalcHEP} we opted for the lowest order $\alpha_s$ evaluation, which is appropriate for a lowest order PDF 
like {\tt CTEQ6L}. The renormalization/factorization scale in {\tt CalcHEP} is set at $\sqrt{\hat{s}}$. This choice of scale results in a somewhat conservative estimate for the event rates.

\begin{figure}
\begin{center}
%
\hskip -15pt
{\epsfig{file=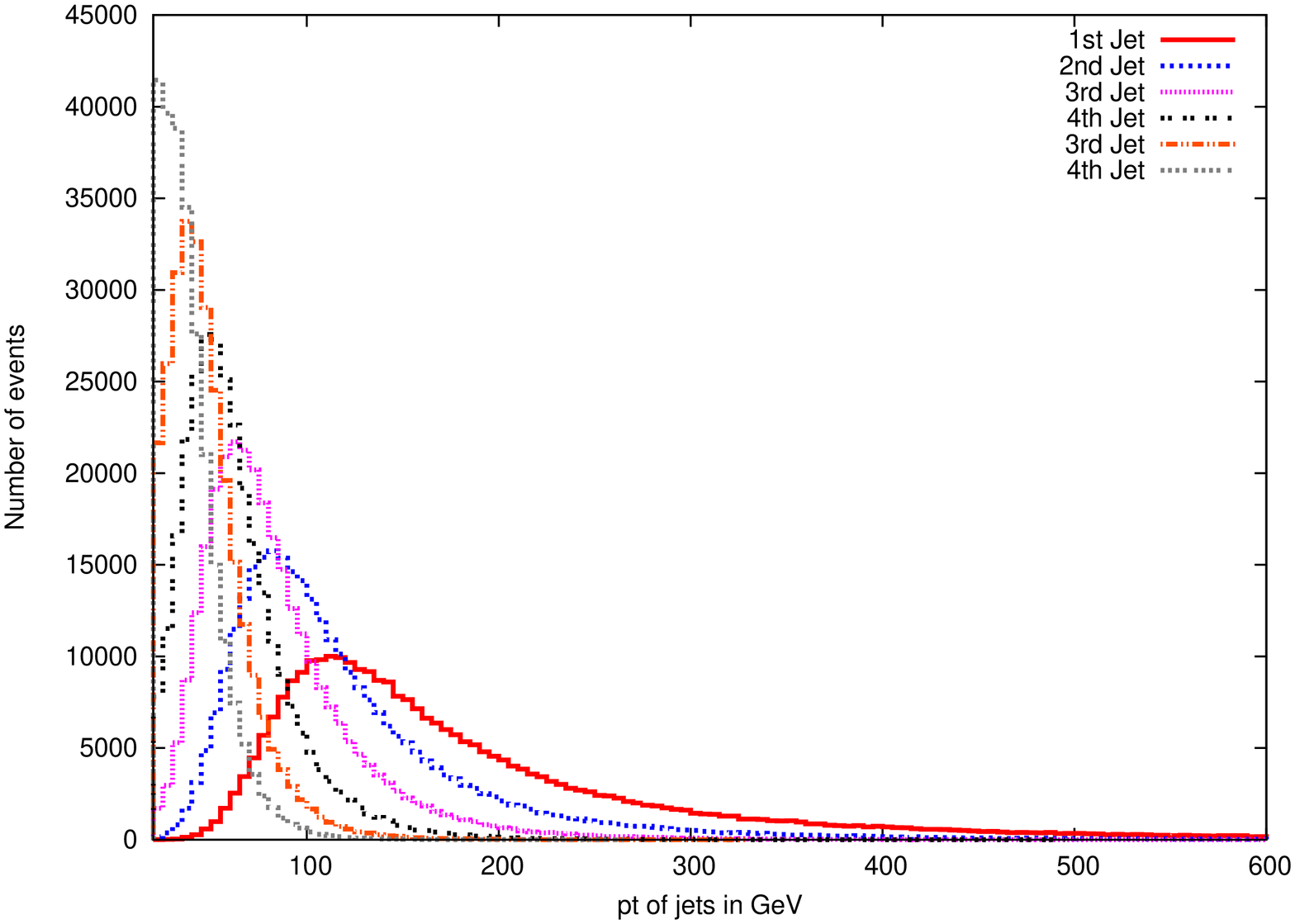,width=7.0 cm,height=8.0cm,angle=-90}}
\hskip -12pt 
{\epsfig{file=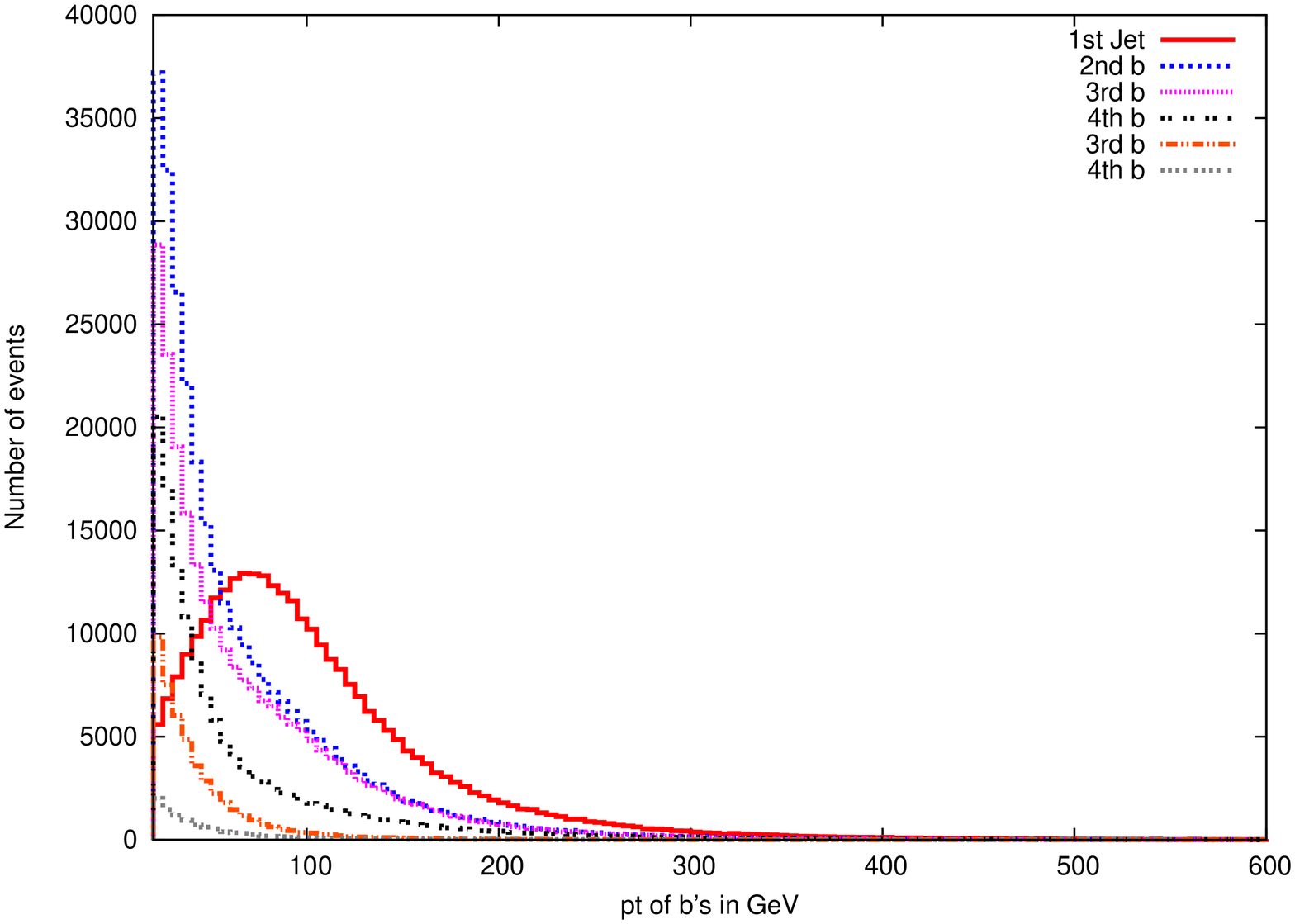,width=7.0cm,height=8.0cm,angle=-90}}
%
\caption{Ordered $p^{jet}_T$ (left) and parton level b $p_T$ distributions in CPV-SUSY scenario for $\tilde{b}_1\tilde{b}^{*}_{1}$}\label{bptdis}
\end{center}
\end{figure}

In the CPX scenario, although $h_1$ decays dominantly into $b\bar{b}$, our simulation reveals that in a fairly large 
fraction of events both the $b$-quarks do not lead to sufficiently hard jets
with reasonable $b$-tagging efficiency. This is because of the lightness of $h_1$ in this scenario.
To illustrate this, we present in Figure \ref{bptdis} the ordered $p_T$ distributions for the four parton-level $b$-quarks in the signal from $\sbot1 \sbotc1$. It is clear from this figure that the 
$b$-quark with the lowest $p_T$ in a given event is
often below 40 GeV or thereabout, which could have ensured a 
moderate tagging efficiency ($\geq$ 50\%).

For hadronic level simulation we have used {\tt PYCELL}, the toy calorimeter simulation provided in
{\tt PYTHIA}, with the following criteria:
\begin{itemize}
  \item the calorimeter coverage is $\rm |\eta| < 4.5$ and the segmentation is
given by $\rm\Delta\eta\times\Delta\phi= 0.09 \times 0.09 $ which resembles
        a generic LHC detector
  \item a cone algorithm with
        $\Delta R = \sqrt{\Delta\eta^{2}+\Delta\phi^{2}} = 0.5$
        has been used for jet finding
  \item $ p_{T,min}^{jet} = 20$ GeV and jets are ordered in $p_{T}$
  \item leptons ($\rm \ell=e,~\mu$) are selected with
        $p_T \ge 20$ GeV and $\rm |\eta| \le 2.5$
  \item no jet should match with a hard lepton in the event

\end{itemize}
From the Figure \ref{bptdis} it is very clear that the parton level distributions
are reflected in the PYCELL level jets. From the $p_T$s of the jets and $b$'s
it is clear that there are two $b$'s or jets having $p_T$ greater than
100 GeV, which are coming from the $\tilde{t}_1$'s. On the other hand
the $b$'s coming from the light Higgs $h$ are of low $p_T$s.

\begin{figure}
\begin{center}
{\epsfig{file=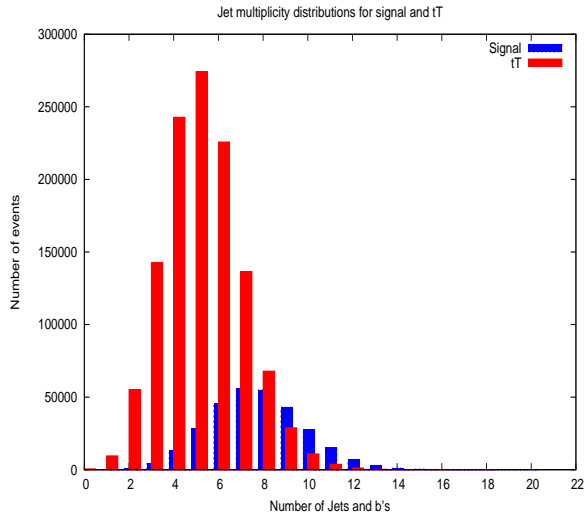,width=7.0 cm,height=8.0cm,angle=-90}}
\caption{Jet multiplicity distributions in CPV-SUSY scenario for
  $\tilde{t}_1\tilde{t}^{*}_{1}$ and $t\bar{t}$}\label{jetm}
\end{center}
\label{fig2}
\end{figure}

\begin{figure}
\begin{center}
%

{\epsfig{file=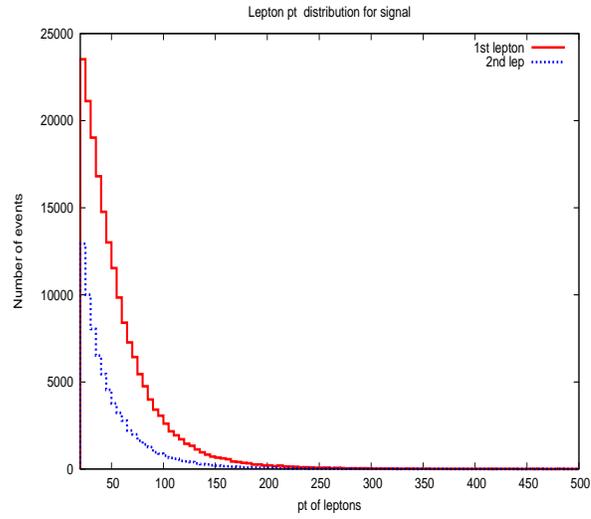,width=7.0 cm,height=8.0cm,angle=-90}}
\caption{Lepton $p_T$ distributions in CPV-SUSY scenario for
  $\tilde{t}_1\tilde{t}^{*}_{1}$ }\label{leppt}
\end{center}
\label{fig3}
\end{figure}

\begin{figure}
\begin{center}
%

{\epsfig{file=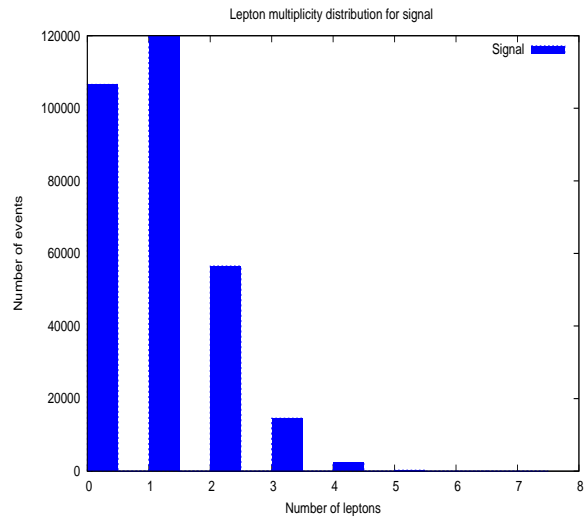,width=7.0 cm,height=8.0cm,angle=-90}}
\caption{Lepton multiplicity distributions in CPV-SUSY scenario for
  $\tilde{t}_1\tilde{t}^{*}_{1}$ }\label{lepm}
\end{center}
\label{fig4}
\end{figure}

From Figure \ref{jetm} we can see that the jet multiplicity for the  $\tilde{t}_1\tilde{t}^{*}_{1}$  is higher than of $t\bar{t}$ which is true for $\tilde{b}_1\tilde{b}^{*}_{1}$ and $\tilde{g}\gluino$ as well. So higher jet multiplicity cut will reduce the $t\bar{t}$ as well as the other Standard Model (SM) backgrounds. Figure \ref{leppt} and \ref{lepm} show the lepton $p_T$ and multiplicity distributions. These figures suggest that we can have some hard leptons in the final states that will also kill the  SM backgrounds; specially with higher multiplicity.

\begin{figure}[hbt]
\begin{center}
%

{\epsfig{file=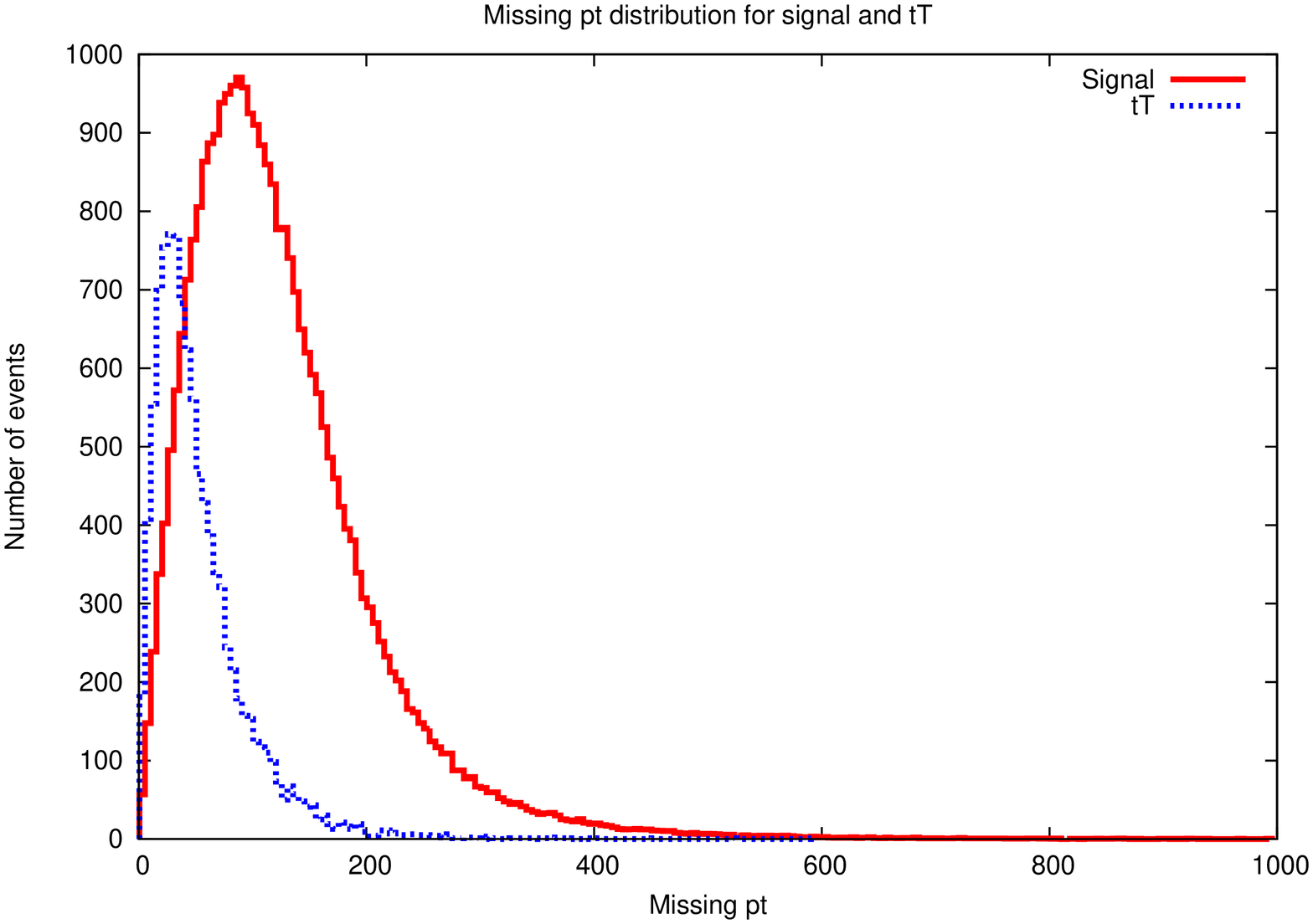,width=7.0 cm,height=8.0cm,angle=-90}}
\caption{Missing $p_T$ distributions in CPV-SUSY scenario for
  $\tilde{b}_1\tilde{b}^{*}_{1}$ and $t\bar{t}$ }\label{mispt}
\end{center}
\label{fig4}
\end{figure}

We also compare the missing $p_T$ distributions of $\tilde{b}_1\tilde{b}^{*}_{1}$ and $t\bar{t}$ in Figure \ref{mispt} which suggests 
that $\ptmiss \geq 100$ GeV will kill the sufficient amount of $t\bar{t}$ events.


The possible backgrounds are very less because of 
higher jet multiplicity and higher number of $b$-jets, leptons 
and missing energy for the signal topologies. But still
for the signal topologies with lower  number of $b$-jets and leptons
can have some Standard model backgrounds. These are $t\bar{t}$, 
$t\bar{t}Z$, $t\bar{t}t\bar{t}$, $t\bar{t}b\bar{b}$.
In particular ISR/FSR can increase the
jet multiplicity of these backgrounds and these should be considered.
 Among these $t\bar{t}t\bar{t}$ cross-sec is very low (of the order few fb), \footnote{Checked with Alpgen \cite{Mangano:2002ea}}.
We calculated other potential backgrounds by CalcHEP-PYTHIA \cite{alex} and AlPGEN-PYTHIA interfacing \cite{Mangano:2002ea}.

\section{Results}

Below we have analysed few signal topologies for the case of bench mark point 1(BP1). Table 
\ref{sigbp1} presents the contributions to the signals for different supersymmetric cascade
decays. In all we have taken nine different signal topologies. Out of these, the first six 
are of higher jet-multiplicities and the remaining topologies are of relatively low jet multiplicities
($n_{jet} \geq 4$). Table  \ref{sigbp1} gives the number for an integrated luminosity of 10 fb$^{-1}$.
From Table \ref{sigbp1} it is clear that for higher jet-multiplicity $\sstop1\sstopc1$ does not 
have much contribution as explained in Section 3. It contributes mostly for the low jet multiplicity 
signals. For the higher jet multiplicity the maximum contribution comes from  $\tilde{g}\tilde{g}$.

The corresponding main SM backgrounds are listed in Table \ref{bgbp1} for an integrated luminosity of 10 fb$^{-1}$.
Comparing Table \ref{sigbp1} and Table \ref{bgbp1} we can get a significance $\geq 10 \sigma$ for almost all the signal 
topologies except signal topology 7; which is of $4.81 \sigma$. Now as we have discussed in Section 3. Out of all the jets
there are two $b-jets$ which are of high $p_T$ coming from the $\sstop1$ decay which are not there in case of the SM backgrounds.
So we demand the first and the second highest jets to have $p_T$ greater than 100 GeV. The event rates with this cuts are given in 
the brackets of Table \ref{sigbp1} and Table \ref{bgbp1}. Implementation of these cuts increases the signal significance by 10-20\%.

\begin{tiny}
\begin{table}
\begin{center}
\begin{tabular}{||c|c|c|c|c||} \hline\hline

No.&Signal topology & $\sbot1\sbotc1$ & $\sstop1\sstopc1$&
$\tilde{g}\tilde{g} $\\
\hline\hline
1&$n_{jet}\ge 8( b-{\rm jet}\ge 3) + \l\ge 2 + \ptmiss \ge 100$&10(5.6)&0.4(0.2)&53(52.8)\\
\hline
2&$n_{jet}\ge 8( b-{\rm jet}\ge 3) + \l\ge 2(\rm{OSD}\ge1) + \ptmiss \ge
100$&7(3.9)&0.4(0.2)&37(36.7)\\
\hline
3&$n_{jet}\ge 8( b-{\rm jet}\ge 3) + \l\ge 2(\rm{SSD}\ge1) + \ptmiss \ge
100$&4(2.2)&0(0)&23(22.1)\\
\hline
4&$n_{jet}\ge 8( b-{\rm jet}\ge 2) + \l\ge 3 + \ptmiss \ge 100$&2(1.1)&0(0)&8(8)\\
\hline
5&$n_{jet}\ge 8( b-{\rm jet}\ge 2) + \l\ge 4 + \ptmiss \ge 100$&0(0)&0(0)&1(0.8)\\
\hline
6&$n_{jet}\ge 8( b-{\rm jet}\ge 4) + \l\ge 2 + \ptmiss \ge 100$&3(1.5)&0(0)&34(33.2)\\
\hline
7&$n_{jet}\ge 4( b-{\rm jet}\ge 3) + \l\ge 1 + \ptmiss \ge 100$&116(63.6)&45(26.2)&283(279.3)\\
\hline
8&$n_{jet}\ge 4( b-{\rm jet}\ge 3) + \l\ge 2(\rm{OSD}\ge1) + \ptmiss \ge 100$&21(9.7)&4(1.9)&54(52.9)\\
\hline
9&$n_{jet}\ge 8( b-{\rm jet}\ge 3)  + \ptmiss \ge 100$&149(96.3)&46(34.2)&499(498)\\

\hline\hline
\end{tabular}
\vspace*{0.0mm}
\caption{Event rates for the CPX point(BP1) of an integrated luminosity of 10 fb$^{-1}$}\label{sigbp1}
\label{tab1}
\end{center}
\end{table}
\begin{table}
\begin{center}
\begin{tabular}{||c|c|c|c|c||} \hline\hline

No.&Signal topology & $t\bar{t}$ & $t\bar{t}Z$&$t\bar{t}b\bar{b} $\\
\hline\hline
1&$n_{jet}\ge 8( b-{\rm jet}\ge 3) + \l\ge 2 + \ptmiss \ge 100$&19(13)&0.33(0.27)&6.1(4.6)\\
\hline
2&$n_{jet}\ge 8( b-{\rm jet}\ge 3) + \l\ge 2(\rm{OSD}\ge1) + \ptmiss \ge
100$&17(12)&0.29(.23)&6.1(4.6)\\
\hline
3&$n_{jet}\ge 8( b-{\rm jet}\ge 3) + \l\ge 2(\rm{SSD}\ge1) + \ptmiss \ge
100$&3(1)&0.05(0.05)&0(0)\\
\hline
4&$n_{jet}\ge 8( b-{\rm jet}\ge 2) + \l\ge 3 + \ptmiss \ge 100$&0(0)&0.27(0.19)&0(0)\\
\hline
5&$n_{jet}\ge 8( b-{\rm jet}\ge 2) + \l\ge 4 + \ptmiss \ge 100$&0(0)&0.0(0.0)&0(0)\\
\hline
6&$n_{jet}\ge 8( b-{\rm jet}\ge 4) + \l\ge 2 + \ptmiss \ge 100$&5(5)&0.08(0.05)&2.6(2.4)\\
\hline
7&$n_{jet}\ge 4( b-{\rm jet}\ge 3) + \l\ge 1 + \ptmiss \ge 100$&1890(953)&22.6(13.21)&297.1	(170.4)\\
\hline
8&$n_{jet}\ge 4( b-{\rm jet}\ge 3) + \l\ge 2(\rm{OSD}\ge1) + \ptmiss \ge 100$&226(101)&2.7(1.4)&34.2(16.6)\\
\hline
9&$n_{jet}\ge 8( b-{\rm jet}\ge 3)  + \ptmiss \ge 100$&1109(784)&13.4(10.5)&252.3(185.6)\\

\hline\hline
\end{tabular}
\vspace*{0.0mm}
\caption{Event rates for the CPX point(BP1) of an integrated luminosity of 10 fb$^{-1}$}\label{bgbp1}
\label{tab1}
\end{center}
\end{table}
\end{tiny}
Next we extend this analysis to the other points of the `LEP-hole'
and show the viability of these signals in other regions of the CPX hole.
 It has already been noted in the literature that
the size and the exact location of the hole in the parameter space
depend on the method of calculating the loop corrections \cite{Lee:2003nta,Frank:2006yh,Hahn:2006np}.
However, the  calculations agree qualitatively and confirm the presence of the
hole. To be specific we have chosen points from  the hole
as presented by \cite{Bechtle:2006iw}.

\begin{table}
\renewcommand\baselinestretch{0.2}
\begin{center}
\begin{tabular}{||c|c|c|c||}
\hline\hline
Parameters&BP2&BP3&BP4\\
\hline\hline
$\tan{\beta}$&4.0&4.0&7.0\\
\hline
$m_{H^\pm}$&140&135&125\\
\hline
$m_{h_1}$ (GeV) &49.45&33.8&40.8\\
\hline\hline
\end{tabular}
\vspace*{0.0mm}
\caption{Benchmark points within the LEP-hole in $m_{h_1}$-$\tan\beta$ plane.}\label{lephole}
\label{tab6}
\end{center}
\end{table}

In Table \ref{lephole} we varied $\tan{\beta}$ and $m_{H^{\pm}}$ keeping the other 
parameters fixed. These correspond to three more different regions of the LEP hole and are
termed as benchmark points 2 -4 (BP2 - BP4), all within the hole. We computed the sensitivity 
for all of these points; both signals and backgrounds. The results are summarised in Tables
\ref{sigbp2}, \ref{bgbp2}, \ref{sigbp3}, \ref{bgbp3}, \ref{sigbp4}, \ref{bgbp4}.  These results suggest
that the LEP-hole can be probed with an integrated luminosity of 5-10 fb$^{-1}$ as like in the benchmark point 1. However, the multilepton channels, e.g., $3\l,\, 4\l$ and like sign di-lepton being almost background free, have $5\sigma$ reach with  very early data of the LHC. So with the early data of LHC every corner of the `hole' is expected to be probed via these cascade decays. For $\sqrt{S}=7$ TeV, the production cross-sections
 are given in Table \ref{tabcross2}. From Table \ref{tabcross2} it is clear that the cross-sections  for  $\sqrt{S}=7$ TeV are reduced by the factor of $\sim 10$, whereas the main background $t\bar{t}$ is reduced by a factor of $\sim 6$. These
 make the reach possible for higher luminosity, though $3\l,\, 4\l$ and like sign di-lepton channels, which are almost background free can be still observed with  the early data of the LHC ( a few fb$^{-1}$).

\begin{tiny}
\begin{table}
\begin{center}
\begin{tabular}{||c|c|c|c|c||} \hline\hline

No.&Signal topology & $\sbot1\sbotc1$ & $\sstop1\sstopc1$&
$\tilde{g}\tilde{g} $\\
\hline\hline
1&$n_{jet}\ge 8( b-{\rm jet}\ge 3) + \l\ge 2 + \ptmiss \ge 100$&14.1(7.8)&0.19(0.10)&58.4(56.5)\\
\hline
2&$n_{jet}\ge 8( b-{\rm jet}\ge 3) + \l\ge 2(\rm{OSD}\ge1) + \ptmiss \ge
100$&10.1(5.6)&0.19(0.10)&40.0(38.7)\\
\hline
3&$n_{jet}\ge 8( b-{\rm jet}\ge 3) + \l\ge 2(\rm{SSD}\ge1) + \ptmiss \ge
100$&5.2(2.8)&0(0)&26.1(25.2)\\
\hline
4&$n_{jet}\ge 8( b-{\rm jet}\ge 2) + \l\ge 3 + \ptmiss \ge 100$&1.8(1.0)&0(0)&9.5(9.1)\\
\hline
5&$n_{jet}\ge 8( b-{\rm jet}\ge 2) + \l\ge 4 + \ptmiss \ge 100$&0(0)&0(0)&0.6(0.6)\\
\hline
6&$n_{jet}\ge 8( b-{\rm jet}\ge 4) + \l\ge 2 + \ptmiss \ge 100$&4.6(2.5)&0.1(0)&36.3(38.7)\\
\hline
7&$n_{jet}\ge 4( b-{\rm jet}\ge 3) + \l\ge 1 + \ptmiss \ge 100$&146.3(78.6)&44.8(23.4)&296.0(289.4)\\
\hline
8&$n_{jet}\ge 4( b-{\rm jet}\ge 3) + \l\ge 2(\rm{OSD}\ge1) + \ptmiss \ge 100$&26.9(12.6)&3.3(1.4)&56.5(54.5)\\
\hline
9&$n_{jet}\ge 8( b-{\rm jet}\ge 3)  + \ptmiss \ge 100$&196.6(123.7)&48.5(34.0)&516.6(510.7)\\

\hline\hline
\end{tabular}
\vspace*{0.0mm}
\caption{Event rates for the CPX point(BP2) of an integrated luminosity of 10 fb$^{-1}$}\label{sigbp2}
\label{tab1}
\end{center}
\end{table}
\begin{table}
\begin{center}
\begin{tabular}{||c|c|c|c|c||} \hline\hline

No.&Signal topology & $t\bar{t}$ & $t\bar{t}Z$&$t\bar{t}b\bar{b} $\\
\hline\hline
1&$n_{jet}\ge 8( b-{\rm jet}\ge 3) + \l\ge 2 + \ptmiss \ge 100$&19(12)&0.33(0.25)&3.4(1.9)\\
\hline
2&$n_{jet}\ge 8( b-{\rm jet}\ge 3) + \l\ge 2(\rm{OSD}\ge1) + \ptmiss \ge
100$&17(10)&0.30(.22)&3.4(1.9)\\
\hline
3&$n_{jet}\ge 8( b-{\rm jet}\ge 3) + \l\ge 2(\rm{SSD}\ge1) + \ptmiss \ge
100$&3(3)&0.07(0.05)&0(0)\\
\hline
4&$n_{jet}\ge 8( b-{\rm jet}\ge 2) + \l\ge 3 + \ptmiss \ge 100$&0(0)&0.17(0.10)&0(0)\\
\hline
5&$n_{jet}\ge 8( b-{\rm jet}\ge 2) + \l\ge 4 + \ptmiss \ge 100$&0(0)&0.0(0.0)&0(0)\\
\hline
6&$n_{jet}\ge 8( b-{\rm jet}\ge 4) + \l\ge 2 + \ptmiss \ge 100$&2(1)&0.12(0.08)&1.9(0.9)\\
\hline
7&$n_{jet}\ge 4( b-{\rm jet}\ge 3) + \l\ge 1 + \ptmiss \ge 100$&1950(952)&23.0(13.4)&314.2(172.6)\\
\hline
8&$n_{jet}\ge 4( b-{\rm jet}\ge 3) + \l\ge 2(\rm{OSD}\ge1) + \ptmiss \ge 100$&226(98)&2.6(1.4)&33.5(15.4)\\
\hline
9&$n_{jet}\ge 8( b-{\rm jet}\ge 3)  + \ptmiss \ge 100$&1056(733)&14.69(11.14)&265.9(187.2)\\

\hline\hline
\end{tabular}
\vspace*{0.0mm}
\caption{Event rates for the CPX point(BP2) of an integrated luminosity of 10 fb$^{-1}$}\label{bgbp2}
\label{tab1}
\end{center}
\end{table}
\end{tiny}

\begin{tiny}
\begin{table}
\begin{center}
\begin{tabular}{||c|c|c|c|c||} \hline\hline

No.&Signal topology & $\sbot1\sbotc1$ & $\sstop1\sstopc1$&
$\tilde{g}\tilde{g} $\\
\hline\hline
1&$n_{jet}\ge 8( b-{\rm jet}\ge 3) + \l\ge 2 + \ptmiss \ge 100$&13.2(7.4)&0.57(0.57)&54.52(53.00)\\
\hline
2&$n_{jet}\ge 8( b-{\rm jet}\ge 3) + \l\ge 2(\rm{OSD}\ge1) + \ptmiss \ge
100$&9.1(5.1)&0.57(0.57)&37.09(36.04)\\
\hline
3&$n_{jet}\ge 8( b-{\rm jet}\ge 3) + \l\ge 2(\rm{SSD}\ge1) + \ptmiss \ge
100$&5.2(2.8)&0(0)&23.18(22.48)\\
\hline
4&$n_{jet}\ge 8( b-{\rm jet}\ge 2) + \l\ge 3 + \ptmiss \ge 100$&1.7(0.9)&0(0)&7.33(6.93)\\
\hline
5&$n_{jet}\ge 8( b-{\rm jet}\ge 2) + \l\ge 4 + \ptmiss \ge 100$&0.1(0)&0(0)&0.82(0.65)\\
\hline
6&$n_{jet}\ge 8( b-{\rm jet}\ge 4) + \l\ge 2 + \ptmiss \ge 100$&4.3(2.2)&0.28(0.28)&34.39(33.39)\\
\hline
7&$n_{jet}\ge 4( b-{\rm jet}\ge 3) + \l\ge 1 + \ptmiss \ge 100$&134.1(72.2)&43.87(20.78)&287.66(282.32)\\
\hline
8&$n_{jet}\ge 4( b-{\rm jet}\ge 3) + \l\ge 2(\rm{OSD}\ge1) + \ptmiss \ge 100$&24.4(11.4)&4.76(2.38)&53.41(51.82)\\
\hline
9&$n_{jet}\ge 8( b-{\rm jet}\ge 3)  + \ptmiss \ge 100$&179.9(114.5)&50.45(33.66)&494.04(488.58)\\

\hline\hline
\end{tabular}
\vspace*{0.0mm}
\caption{Event rates for the CPX point(BP3) of an integrated luminosity of 10 fb$^{-1}$}\label{sigbp3}
\label{tab1}
\end{center}
\end{table}

\begin{table}
\begin{center}
\begin{tabular}{||c|c|c|c|c||} \hline\hline

No.&Signal topology & $t\bar{t}$ & $t\bar{t}Z$&$t\bar{t}b\bar{b} $\\
\hline\hline
1&$n_{jet}\ge 8( b-{\rm jet}\ge 3) + \l\ge 2 + \ptmiss \ge 100$&9(5)&0.32(0.21)&6.8(3.9)\\
\hline
2&$n_{jet}\ge 8( b-{\rm jet}\ge 3) + \l\ge 2(\rm{OSD}\ge1) + \ptmiss \ge
100$&9(5)&0.29(.18)&6.5(3.7)\\
\hline
3&$n_{jet}\ge 8( b-{\rm jet}\ge 3) + \l\ge 2(\rm{SSD}\ge1) + \ptmiss \ge
100$&3(3)&0.07(0.03)&0.2(0.2)\\
\hline
4&$n_{jet}\ge 8( b-{\rm jet}\ge 2) + \l\ge 3 + \ptmiss \ge 100$&0(0)&0.27(0.20)&0(0)\\
\hline
5&$n_{jet}\ge 8( b-{\rm jet}\ge 2) + \l\ge 4 + \ptmiss \ge 100$&0(0)&0.0(0.0)&0(0)\\
\hline
6&$n_{jet}\ge 8( b-{\rm jet}\ge 4) + \l\ge 2 + \ptmiss \ge 100$&1(0)&0.10(0.05)&2.9(1.7)\\
\hline
7&$n_{jet}\ge 4( b-{\rm jet}\ge 3) + \l\ge 1 + \ptmiss \ge 100$&1250(491)&22.05(12.43)&313.8(174.5)\\
\hline
8&$n_{jet}\ge 4( b-{\rm jet}\ge 3) + \l\ge 2(\rm{OSD}\ge1) + \ptmiss \ge 100$&130(41)&2.90(1.32)&38.5(17.1)\\
\hline
9&$n_{jet}\ge 8( b-{\rm jet}\ge 3)  + \ptmiss \ge 100$&473(266)&13.99(10.44)&254.9(178.2)\\

\hline\hline
\end{tabular}
\vspace*{0.0mm}
\caption{Event rates for the CPX point(BP3) of an integrated luminosity of 10 fb$^{-1}$}\label{bgbp3}
\label{tab1}
\end{center}
\end{table}
\end{tiny}

\begin{tiny}
\begin{table}
\begin{center}
\begin{tabular}{||c|c|c|c|c||} \hline\hline

No.&Signal topology & $\sbot1\sbotc1$ & $\sstop1\sstopc1$&
$\tilde{g}\tilde{g} $\\
\hline\hline
1&$n_{jet}\ge 8( b-{\rm jet}\ge 3) + \l\ge 2 + \ptmiss \ge 100$&3.39(2.09)&0.19(0.09)&50.53(49.24)\\
\hline
2&$n_{jet}\ge 8( b-{\rm jet}\ge 3) + \l\ge 2(\rm{OSD}\ge1) + \ptmiss \ge
100$&2.36(1.44)&0.09(0.00)&33.86(33.04)\\
\hline
3&$n_{jet}\ge 8( b-{\rm jet}\ge 3) + \l\ge 2(\rm{SSD}\ge1) + \ptmiss \ge
100$&1.28(0.80)&0(0)&22.18(21.48)\\
\hline
4&$n_{jet}\ge 8( b-{\rm jet}\ge 2) + \l\ge 3 + \ptmiss \ge 100$&0.74(0.51)&0(0)&7.28(6.98)\\
\hline
5&$n_{jet}\ge 8( b-{\rm jet}\ge 2) + \l\ge 4 + \ptmiss \ge 100$&0.03(0.03)&0(0)&0.47(0.47)\\
\hline
6&$n_{jet}\ge 8( b-{\rm jet}\ge 4) + \l\ge 2 + \ptmiss \ge 100$&0.63(0.37)&0(0)&29.40(28.76)\\
\hline
7&$n_{jet}\ge 4( b-{\rm jet}\ge 3) + \l\ge 1 + \ptmiss \ge 100$&66.47(38.48)&41.47(21.74)&314.14(307.86)\\
\hline
8&$n_{jet}\ge 4( b-{\rm jet}\ge 3) + \l\ge 2(\rm{OSD}\ge1) + \ptmiss \ge 100$&10.18(5.18)&4.01(2.00)&59.22(57.16)\\
\hline
9&$n_{jet}\ge 8( b-{\rm jet}\ge 3)  + \ptmiss \ge 100$&77.26(52.73)&47.11(34.01)&521.51(515.17)\\

\hline\hline
\end{tabular}
\vspace*{0.0mm}
\caption{Event rates for the CPX point(BP4) of an integrated luminosity of 10 fb$^{-1}$}\label{sigbp4}
\label{tab1}
\end{center}
\end{table}

\begin{table}
\begin{center}
\begin{tabular}{||c|c|c|c|c||} \hline\hline

No.&Signal topology & $t\bar{t}$ & $t\bar{t}Z$&$t\bar{t}b\bar{b} $\\
\hline\hline
1&$n_{jet}\ge 8( b-{\rm jet}\ge 3) + \l\ge 2 + \ptmiss \ge 100$&9(4)&0.29(0.23)&3.7(3.1)\\
\hline
2&$n_{jet}\ge 8( b-{\rm jet}\ge 3) + \l\ge 2(\rm{OSD}\ge1) + \ptmiss \ge
100$&9(4)&0.25(.21)&2.9(2.4)\\
\hline
3&$n_{jet}\ge 8( b-{\rm jet}\ge 3) + \l\ge 2(\rm{SSD}\ge1) + \ptmiss \ge
100$&3(3)&0.07(0.05)&0.7(0.7)\\
\hline
4&$n_{jet}\ge 8( b-{\rm jet}\ge 2) + \l\ge 3 + \ptmiss \ge 100$&0(0)&0.18(0.10)&0(0)\\
\hline
5&$n_{jet}\ge 8( b-{\rm jet}\ge 2) + \l\ge 4 + \ptmiss \ge 100$&0(0)&0.0(0.0)&0(0)\\
\hline
6&$n_{jet}\ge 8( b-{\rm jet}\ge 4) + \l\ge 2 + \ptmiss \ge 100$&1(1)&0.09(0.07)&0.9(0.7)\\
\hline
7&$n_{jet}\ge 4( b-{\rm jet}\ge 3) + \l\ge 1 + \ptmiss \ge 100$&1049(420)&21.43(12.58)&310.6(168.2)\\
\hline
8&$n_{jet}\ge 4( b-{\rm jet}\ge 3) + \l\ge 2(\rm{OSD}\ge1) + \ptmiss \ge 100$&105(30)&2.42(1.30)&38.6(16.1)\\
\hline
9&$n_{jet}\ge 8( b-{\rm jet}\ge 3)  + \ptmiss \ge 100$&412(244)&13.70(10.15)&258.8(184.1)\\

\hline\hline
\end{tabular}
\vspace*{0.0mm}
\caption{Event rates for the CPX point(BP4) of an integrated luminosity of 10 fb$^{-1}$}\label{bgbp4}
\label{tab1}
\end{center}
\end{table}
\end{tiny}

\begin{table}
\begin{center}
\begin{tabular}{||c|c|c|c||}
\hline
 $\sigma_{\sstop1 \sstopc1 }$  &  $\sigma_{\sbot1 \sbotc1 }$  & $\sigma_{\sbot2 \sbotc2 }$ &  $\sigma_{ \tilde{g}\tilde{g}}$ \\
\hline
288.7& 21.9 & 10.6 &0.7\\
\hline
\end{tabular}
\caption{Production cross sections (in fb) at lowest-order 
computed with {\tt CalcHEP} interfaced with CPsuperH
 for different processes at the LHC in the
CPX scenario and for the spectrum of Table 1 for $\sqrt{s}=7$ TeV. 
 {\tt CTEQ6L} parton distribution functions are used and
 the renormalization/factorization scale is set to  $\sqrt{\hat{s}}$.}\label{tabcross2}
\end{center}
\end{table}

\begin{figure}
\begin{center}
%

{\epsfig{file=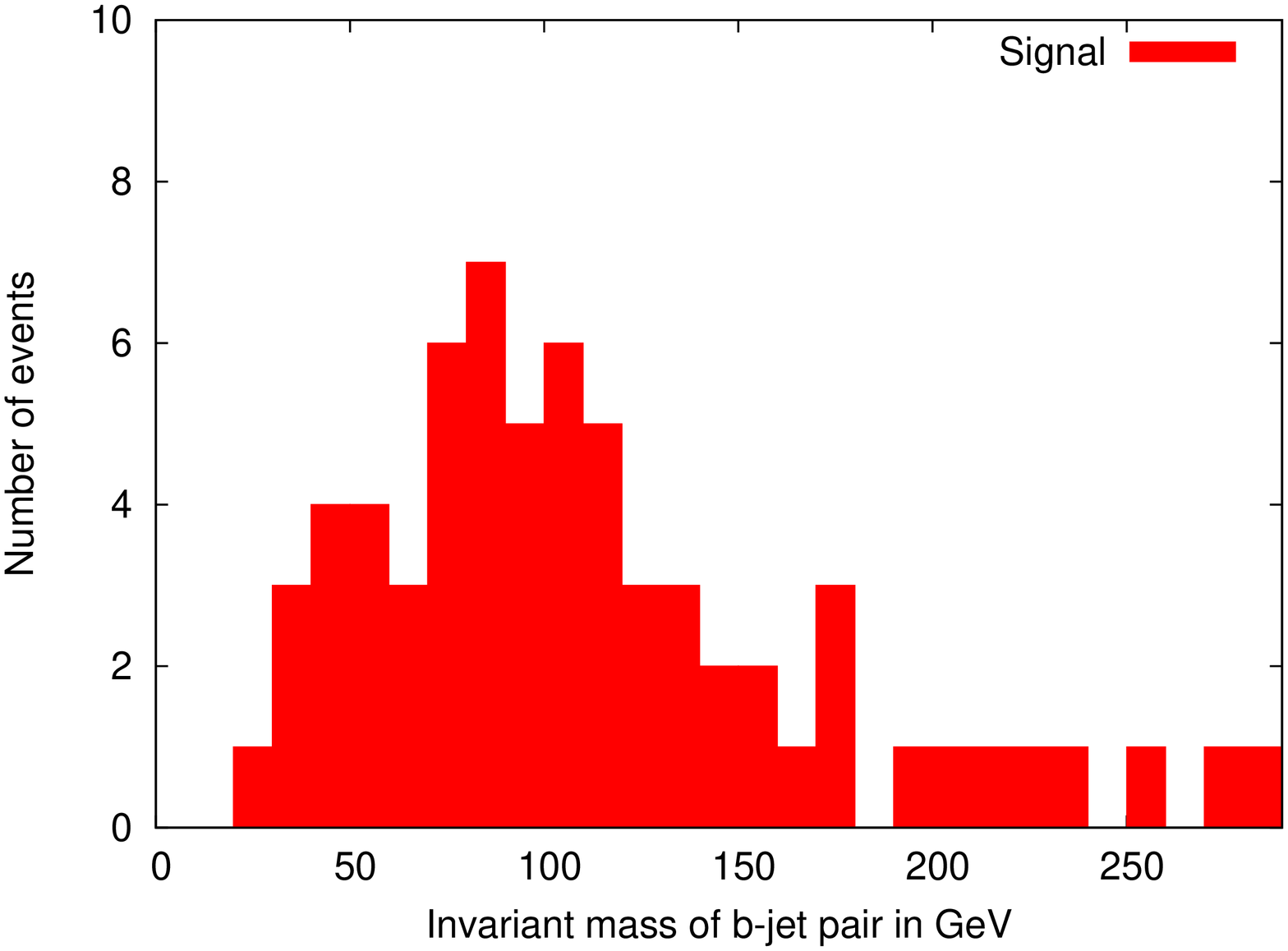,width=7.0 cm,height=8.0cm,angle=-90}}
\caption{Invariant mass of $b$-jet pair for benchmark point 1
}\label{invm}
\end{center}
\end{figure}

On top of this multi-channel analysis we try get
the invariant mass peak for the light Higgs boson. In that case
it would of great help in probing the CP-violating light Higgs
boson as well as killing all the backgrounds which include the 
model background as well. Among the all different
signal topologies we have taken $n-jet \geq 6\, (3\, b-jets)\, + \,4 \,\rm{leptons}\, +\, \ptmiss \geq 150 \,\rm{GeV}\, + \,m_{eff}\geq 100\, \rm{GeV}$ which has the best reach. Here we define $m_{eff}$ as scalar sum of missing-$p_T$, jet-$p_T$s and lepton-$p_T$s. Figure \ref{invm} shows the invariant mass of two b-jets
in this final state topology for benchmark point 1. The figure clearly shows a 
peak around light Higgs boson mass which is around 40 GeV. On top of this at 
integrated luminosity of 10 fb$^{-1}$ no $t\bar{t}$ event passes the above
signal criteria to contribute in the invariant mass distribution as backgrounds.
The mass peak also very useful to kill all the other model backgrounds.
 Thus, reconstruction of the light Higgs boson mass peak could 
be an indicator of the CP-violating Higgs discovery.

\begin{figure}
\begin{center}
%
\hskip -15pt
{\epsfig{file=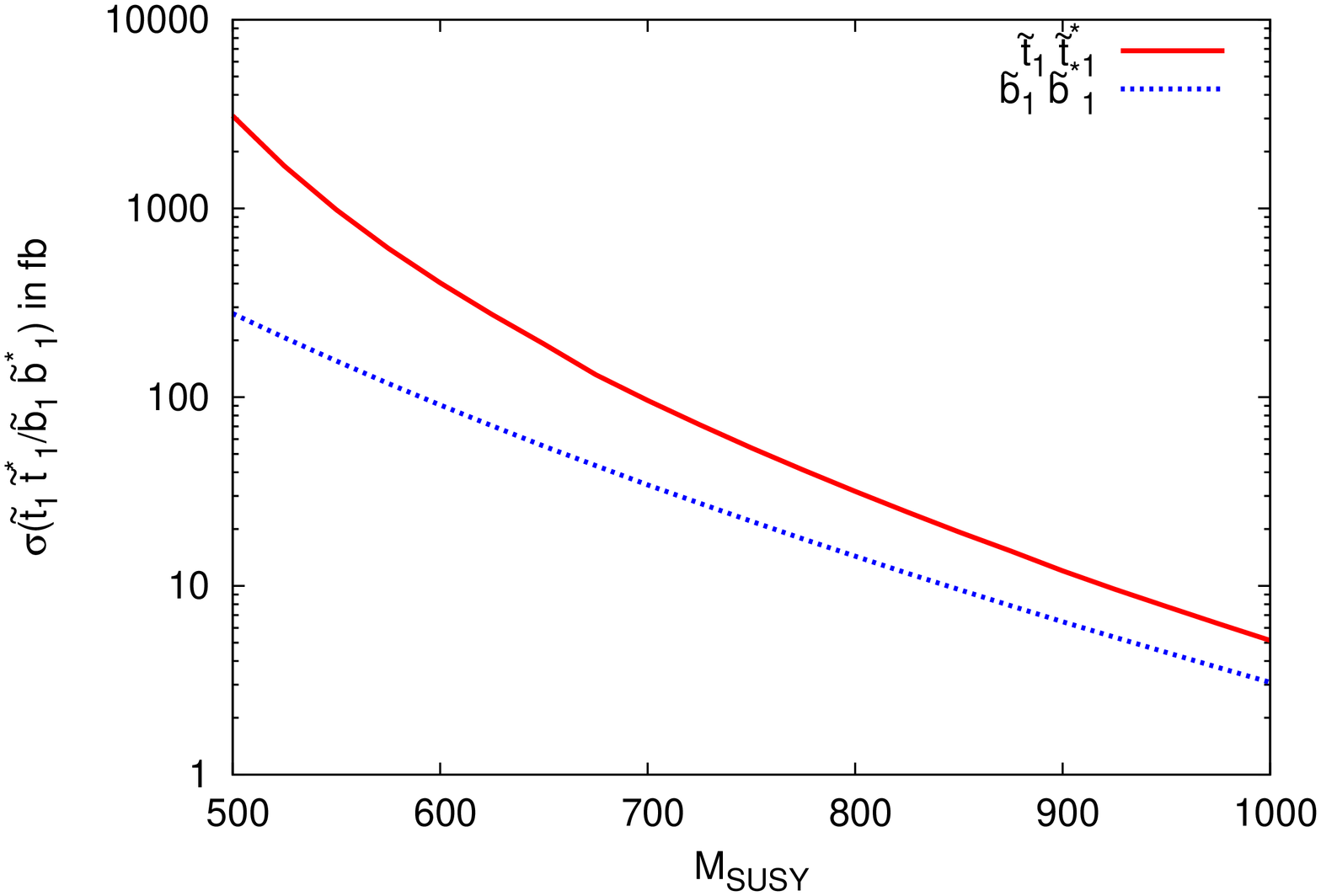,width=7.0 cm,height=8.0cm,angle=-90}}
\hskip -12pt 
{\epsfig{file=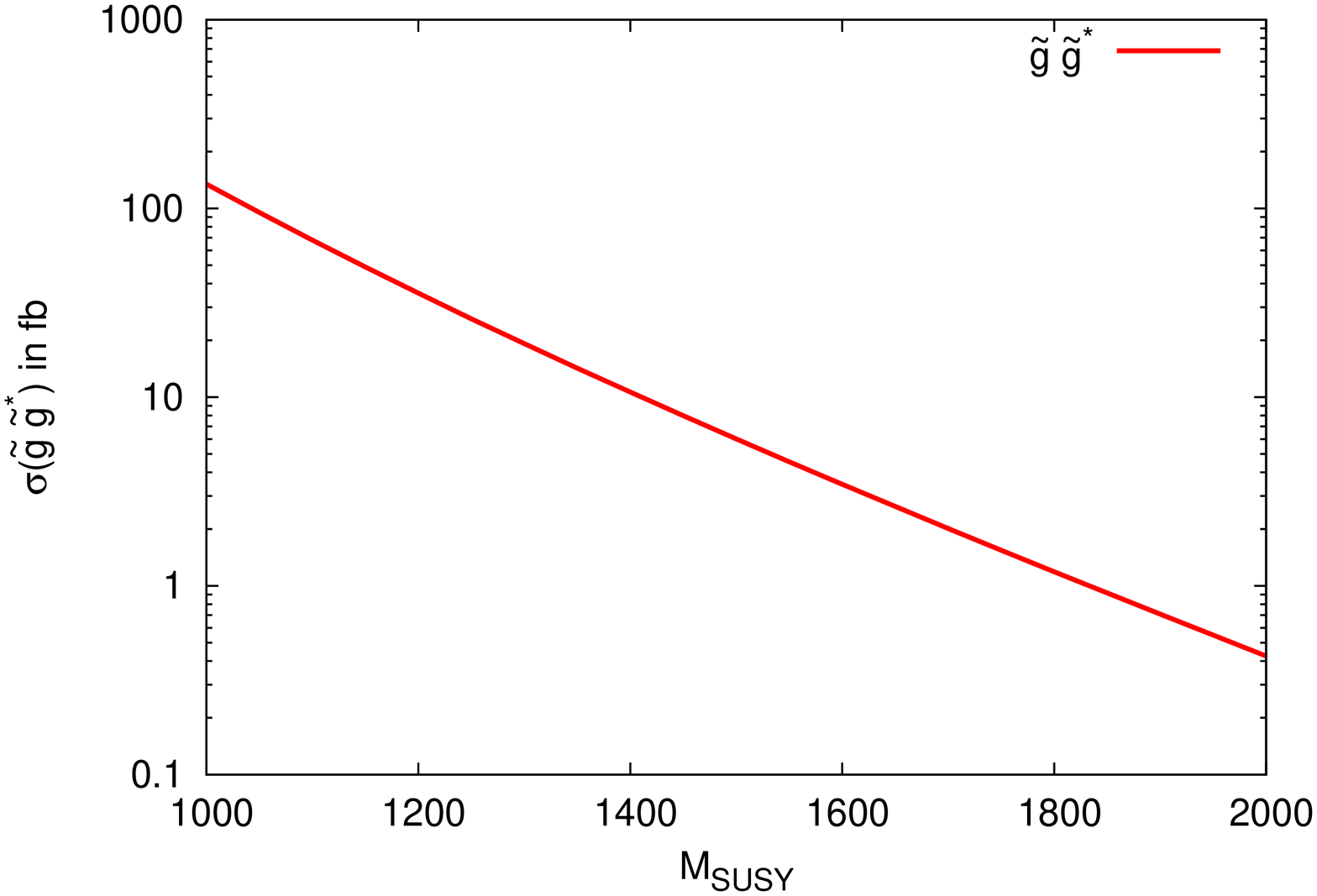,width=7.0cm,height=8.0cm,angle=-90}}
%
\caption{Cross-sec variation of with $M_{\rm SUSY}$ for $\tilde{t}_1\tilde{t}^{*}_{1}$, $\tilde{b}_1\tilde{b}^{*}$ (left) and $\tilde{g}\tilde{g}$ (right)}\label{prodcpx}
\end{center}
\end{figure}

The effective masses and couplings at one loop
depend on the soft masses, i.e., $M_{\rm SUSY}$, that contribute
in the loop. Thus, the $Z-Z-h_1$ coupling as well the shape
of the 'LEP hole' change with the variation of $M_{\rm SUSY}$.
In principle there could be some regions where
$m_{h_1}$ is still light ($\le 60$ GeV) but ruled out by LEP 
because of the $Z-Z-h_1$ coupling is large.
From \cite{Abb} we see that when $M_{\rm SUSY}$ increases the
 'LEP hole' almost vanishes, as the mixing term in the Higgs 
mass matrix, i.e., $M_{\rm SP}\simeq \frac{\mu^2A}{M_{\rm SUSY}}$ goes
to zero. In page 62 and Figure 28 of \cite{Abb} describes two 
different scenarios corresponding to $M_{\rm SUSY}=1$ TeV as given below.
\begin{enumerate}
\item{\bf CPX1.0:}

$\mu=4M_{\rm SUSY},\quad\quad |A|=2 M_{\rm SUSY},\quad\quad |M_3|=2 M_{\rm SUSY}$.

For this case with $M_{\rm SUSY}=1 \rm{TeV}$ the hole is still there near $m_{h_1}=30-60$ GeV.

\item{\bf CPX0.5:}

Where $M_{\rm SUSY}=1$ TeV for third generation squark masses
but all the other parameters kept in the as normal CPX, i.e. corresponding to
$M_{\rm SUSY}=500$ GeV case as before. For this case the 'hole' is
 shifted to $m_{h_1}\ge 75$ GeV.
\end{enumerate}
Figure \ref{prodcpx} describe the variation of the production
cross-section of the $\tilde{t}_1\tilde{t}^*_1$, $\tilde{b}_1\tilde{b}^*_1$
and $\tilde{g}\tilde{g}$ with $M_{\rm SUSY}$ respectively.
From the figures it is very clear that the cross-sections
drops down very fast as $M_{\rm SUSY}$ increases. Thus the
 cross-sections for the {\bf CPX0.5}, {\bf CPX1.0}
decrease, resulting lowering of events for the final states.
In Table \ref{cpx0.5} and Table \ref{cpx1.0} present the 
number of events for the {\bf CPX0.5}, {\bf CPX1.0} for the 
integrated luminosity of 10 fb$^{-1}$. The numbers suggest that
for the {\bf CPX0.5} we still get 5$\sigma$ signal significance 
over the SM backgrounds for most of the signal topologies. This
is because the gluino contribution is still large, 
as in {\bf CPX0.5} the $|m_{\tilde{g}}|=1 \rm{TeV}$; which is same 
as the normal CPX scenario. Where as for {\bf CPX1.0} none of the 
final states get $5\sigma$ significance for 10 fb$^{-1}$ of 
integrated luminosity. The signal 8, which is having best significance
over the background, will require 128 fb$^{-1}$ of integrated 
luminosity for the $5\sigma$ significance over the SM backgrounds. 

\begin{tiny}
\begin{table}
\begin{center}
\begin{tabular}{||c|c|c|c|c||} \hline\hline

No.&Signal topology &$\sbot1\sbotc1$ & $\sstop1\sstopc1$&
$\tilde{g}\tilde{g} $\\
\hline\hline
1&$n_{jet}\ge 8( b-{\rm jet}\ge 3) + \l\ge 2 + \ptmiss \ge 100$&0.18(0.18)&0.06(0.05)&34.11(32.75)\\
\hline
2&$n_{jet}\ge 8( b-{\rm jet}\ge 3) + \l\ge 2(\rm{OSD}\ge1) + \ptmiss \ge
100$&0.15(0.14)&0.06(0.05)&23.18(22.51)\\
\hline
3&$n_{jet}\ge 8( b-{\rm jet}\ge 3) + \l\ge 2(\rm{SSD}\ge1) + \ptmiss \ge
100$&0.06(0.06)&0.01(0.01)&13.89(13.07)\\
\hline
4&$n_{jet}\ge 8( b-{\rm jet}\ge 2) + \l\ge 3 + \ptmiss \ge 100$&0.04(0.04)&0(0)&4.18(3.91)\\
\hline
5&$n_{jet}\ge 8( b-{\rm jet}\ge 2) + \l\ge 4 + \ptmiss \ge 100$&0.01(0.01)&0(0)&0.40(0.40)\\
\hline
6&$n_{jet}\ge 8( b-{\rm jet}\ge 4) + \l\ge 2 + \ptmiss \ge 100$&0.06(0.06)&0.04(0.03)&20.09(19.01)\\
\hline
7&$n_{jet}\ge 4( b-{\rm jet}\ge 3) + \l\ge 1 + \ptmiss \ge 100$&2.43(2.29)&3.24(2.96)&238.88(226.20)\\
\hline
8&$n_{jet}\ge 4( b-{\rm jet}\ge 3) + \l\ge 2(\rm{OSD}\ge1) + \ptmiss \ge 100$&0.35(0.31)&0.36(0.31)&40.85(37.34)\\
\hline
9&$n_{jet}\ge 8( b-{\rm jet}\ge 3)  + \ptmiss \ge 100$&3.33(3.25)&4.16(4.08)&396.06(384.37)\\
\hline\hline
\end{tabular}
\vspace*{0.0mm}
\caption{Event rates for the CPX0.5 of an integrated luminosity of 10 fb$^{-1}$}\label{cpx0.5}
\end{center}
\end{table}
\end{tiny}

\begin{tiny}
\begin{table}
\begin{center}
\begin{tabular}{||c|c|c|c|c||} \hline\hline

No.&Signal topology &$\sbot1\sbotc1$ & $\sstop1\sstopc1$&
$\tilde{g}\tilde{g} $\\
\hline\hline
1&$n_{jet}\ge 8( b-{\rm jet}\ge 3) + \l\ge 2 + \ptmiss \ge 100$&0.16(0.15)&0.12(0.11)&0.25(0.25)\\
\hline
2&$n_{jet}\ge 8( b-{\rm jet}\ge 3) + \l\ge 2(\rm{OSD}\ge1) + \ptmiss \ge
100$&0.11(0.11)&0.09(0.09)&0.17(0.17)\\
\hline
3&$n_{jet}\ge 8( b-{\rm jet}\ge 3) + \l\ge 2(\rm{SSD}\ge1) + \ptmiss \ge
100$&0.06(0.05)&0.03(0.02)&0.12(0.12)\\
\hline
4&$n_{jet}\ge 8( b-{\rm jet}\ge 2) + \l\ge 3 + \ptmiss \ge 100$&0.03(0.03)&0(0)&0.04(0.04)\\
\hline
5&$n_{jet}\ge 8( b-{\rm jet}\ge 2) + \l\ge 4 + \ptmiss \ge 100$&0.01(0.01)&0(0)&0.01(0.01)\\
\hline
6&$n_{jet}\ge 8( b-{\rm jet}\ge 4) + \l\ge 2 + \ptmiss \ge 100$&0.05(0.05)&0.07(0.06)&0.17(0.17)\\
\hline
7&$n_{jet}\ge 4( b-{\rm jet}\ge 3) + \l\ge 1 + \ptmiss \ge 100$&2.54(2.40)&6.32(5.64)&1.33(1.33)\\
\hline
8&$n_{jet}\ge 4( b-{\rm jet}\ge 3) + \l\ge 2(\rm{OSD}\ge1) + \ptmiss \ge 100$&0.43(0.38)&0.56(0.43)&0.23(0.23)\\
\hline
9&$n_{jet}\ge 8( b-{\rm jet}\ge 3)  + \ptmiss \ge 100$&3.68(3.58)&9.48(9.12)&2.35(2.35)\\
\hline\hline
\end{tabular}
\vspace*{0.0mm}
\caption{Event rates for the CPX1.0 of an integrated luminosity of 10 fb$^{-1}$}\label{cpx1.0}
\end{center}
\end{table}
\end{tiny}

In our choice of parameter points, we have
$M_1=100$ GeV, $M_2=200$ GeV and $M_{\rm SUSY}=500$ GeV
for all the first four benchmark points. In those cases
 stop has two decay modes; $\tilde{t}_1\to t \ntrl1$
and $\tilde{t}_1\to b \chp1$.
But as we increase $M_{\rm SUSY}$ to higher values 
other decay modes of stop could also be open. In principle 
stop can decay to $t \ntrl2$ (in the case of {\bf CPX0.5}, {\bf CPX1.0})
as shown below. The $\ntrl2$ thus produced, can have two-body
or three body decays depending on the parameter points 
which will enrich the final state at the end
\begin{eqnarray}\nonumber
\tilde{t}_1\to t \ntrl2\to t ll \ntrl1\\\nonumber
     \to t Z/h \ntrl1\nonumber
\end{eqnarray}.
Along with the above mentioned decay modes, we could have
the following decay and also to higher neutralino modes.
\begin{eqnarray}\nonumber
\tilde{t}_1\to b \chp2 \to b H^{+} \ntrl1\\\nonumber
     \to b W^+ \ntrl1\nonumber
\end{eqnarray}
But, in the CPX scenario, $\mu=4M_{\rm SUSY}$, which makes the higher
charginos and neutralinos as Higgsino type, thus much heavier than the 
corresponding stop mass eigen states. This results in
making these decay modes kinematically disallowed.

\section{Summary and Conclusions}

We have explored supersymmetric cascade decays in the context of CP-violating MSSM
and important for the CPX scenario. This cascade decay analysis can also probe
the `LEP-hole' in the CPX scenario with the early data from LHC. 
These final states in the CPX scenario are a consequence 
of low mass of the lightest Higgs boson (as light as 30 GeV). The invariant
 mass distribution of these $b$-jets also peaks around the lightest neutral Higgs boson. This kills all the model and other backgrounds. 
Moreover, multi-lepton final states ($3\l,\, 4\l$ and like sign di-lepton)
are easy to detect as they are almost background free; thus have $5\sigma$ 
reach for the corresponding signals with very early data of LHC for both 14 TeV and 7 TeV center of mass energy. We have also studied the range of sensitivity
 for $M_{\rm SUSY}=1$ TeV for {\bf CPX0.5} and {\bf CPX1.0} in this context.
We found that though $5\sigma$ reach is possible for {\bf CPX0.5} with 10 fb$^{-1}$ of data, for {\bf CPX1.0} one needs to go beyond 100 fb$^{-1}$ data. Also heavier stop in these cases can lead to richer final states through its decay
to Higgs(es) or leptons. Finally the supersymmetric cascades under CP-violating scenario are very different from the CP-conserving case because of the possible non-trivial decay modes in the former case.

\vskip 25pt
\noindent
{\bf Acknowledgments:} I would like to thank 
Prof. Michael Peskin for pointing out to this problem
of third generation supersymmetric cascade decays in the
CP-violating scenario. I would also like to thank
Prof. Amitava Datta, Prof. Manuel Drees, Prof. Biswarup Mukhopadhyaya and Dr. AseshKrishna Datta
for the useful discussions. I also thank Ms. Nabanita Bhattacharyya
for the discussion regarding $b$-tagging. Computational work for this study was partially carried out in the cluster
computing facility at Harish-Chandra Research Institute (HRI)
(http://cluster.mri.ernet.in).

\vskip 45pt


\end{document}